\documentclass{emulateapj}  

\begin{document}

\title{PREGALACTIC LiBeB PRODUCTION BY SUPERNOVA COSMIC RAYS}

\author{MOTOHIKO KUSAKABE\altaffilmark{1}}
\affil{Department of Astronomy, School of Science, University of
Tokyo,  Hongo, Bunkyo-ku, Tokyo 113-0033, Japan \\
Division of Theoretical Astronomy, National Astronomical
Observatory of Japan, Mitaka, Tokyo
  181-8588, Japan\\
{\tt kusakabe@th.nao.ac.jp}}

\altaffiltext{1}{Research Fellow of the Japan Society for the Promotion
of Science.}

\begin{abstract}
I calculate the evolution of Be and B abundances produced by cosmic
 rays generated by massive stars in the pregalactic phase of the
 universe.  The inputs for calculation, i.e. the star formation rate
 and the nuclear abundances of cosmic rays, which I assume to be the
 same as those of the ISM, are taken from the results of a detailed
 cosmic chemical evolution model with its parameters best fitted from
 several items of observational information including an early reionization of
 the IGM by $z\sim 15$.  I found that when the $^6$Li plateau abundance
 observed in metal-poor halo stars originated in the pregalactic
 cosmological cosmic ray nucleosynthesis, Be and B simultaneously
 produced with $^6$Li amount to the lowest levels ever detected in
 metal-poor halo stars.  It is desirable to observe Be and B abundances
 in metal-poor halo stars with [Fe/H]$\leq -3$ in order to elucidate the
 possibility of early $^6$LiBeB production by pregalactic supernova
 cosmic ray nucleosynthesis.
\end{abstract}

\keywords{cosmic rays ---  cosmology: theory ---  nuclear reactions,
nucleosynthesis, abundances --- stars: abundances --- stars: Population
II --- supernovae: general}

\section{INTRODUCTION}

The lithium abundances observed in metal-poor halo stars (MPHSs) show a
plateau as a function of metallicity 
~\citep{spi1982,rya2000,mel2004,asp2006,bon2007,shi2007} at $^7$Li/H$=1-2$
$\times 10^{-10}$.  The
prediction by the standard big bang nucleosynthesis (BBN) model of
$^7$Li abundance which is the main lithium isotope observed in MPHSs, however, indicates
a factor of $2-4$ larger value, when the baryon-to-photon ratio deduced
from parameter fits to the temperature fluctuations of cosmic microwave
background (CMB) radiation measured with Wilkinson Microwave Anisotropy Probe
(WMAP)~\citep{spe2003,spe2007} is
used.  For example, \citet{coc2004} derived $^7$Li/H=$(4.15^{+0.49}_{-0.45})\times 10^{-10}$
with the baryon-to-photon ratio $\eta=(6.14\pm 0.25)\times 10^{-10}$.
This discrepancy between the observations and the BBN+CMB prediction of
$^7$Li abundance is a problem, which indicates some destruction
process of $^7$Li.  Recently, a complex but consistent theory is
suggested by~\citet{pia2006}.  In their theory, an extremely high
efficiency of engulfment of baryons in a first generation of stars
results in a half or one third of destructions of D and Li isotopes.
Population II (Pop II) stars are made of a mixture of ejecta of
supernovae (SNe) of the
first stars and unprocessed material of the primordial composition, and
experience a depletion of lithium isotopes in their atmospheres, which
are observed.  There are two important conditions.  The ejecta of SNe of the first stars needs to mix with the pure BBN-composition
matter at $2.5\leq$ [Fe/H]\footnote{[A/H]=log(A/H)$-$log(A/H)$_{\sun}$} in order to form the lithium plateau.  An
infall of the intergalactic medium (IGM) needs to proceed after the formation
of Pop II stars in order to be consistent with observations of deuterium
abundance.

Recent spectroscopic observations of MPHSs also provide abundances of
$^6$Li isotope.  They indicate a likely primordial plateau abundance,
similar to the well known $^7$Li plateau, of
$^6$Li/H=$6\times10^{-12}$, which is about 1000 times as large as the
BBN prediction.  Since the standard Galactic
cosmic ray (CR) nucleosynthesis models predict negligible amounts of $^6$Li
abundance with respect to the observed plateau level at [Fe/H] $<-2$~\citep[e.g.][]{pra2006}, this plateau causes another problem, which indicates some production process of $^6$Li.

Several candidates for early $^6$Li production mechanisms have been suggested.  The non-thermal nuclear reactions triggered by the decay of
long-lived particles is one possibility of the non-standard
process~\citep{jed2000,jed2004a,jed2004b,jed2006,kawasaki2005,kus2006,cum2007}.
\citet{pos2006} suggested the exotic nuclear reaction of
$^4$He$_X$($d$,$X^-$)$^6$Li to make abundant $^6$Li, where $X^-$ is a negatively charged massive
particle assumed to decay in the early universe, and $^4$He$_X$ is the
state that has $^4$He bound to $X^-$.
\citet{suz2002} suggested an $\alpha+\alpha$ fusion reaction with $\alpha$
particles accelerated by hierarchical structure formation shocks, thought
to have been operative at the Galaxy formation epoch.

As a possibility, \citet*{rol2005} have calculated the $^6$Li
production by an initial burst of cosmological cosmic rays (CCRs) to
show that this process through $\alpha+\alpha$ fusion can account for the $^6$Li plateau without
overproduction of $^7$Li.  \citet{rol2006} applied the CCR
nucleosynthesis to a well grounded detailed model.  They
derived the total CCR energy as a function of redshift from a star
formation rates (SFRs) in models of cosmic chemical evolution of
\citet{dai2006}, which are made to reproduce the observed cosmic SFR, SN
II rate, the present fraction of baryons in structures, that in stars,
the evolution of the metal content in the interstellar medium (ISM) and IGM, and early
reionization of the IGM.  As a result they found that the pregalactic
production of the $^6$Li in the IGM via Population III (Pop III) stars can account
for the $^6$Li plateau without overproduction of $^7$Li.

The CR nucleosynthesis also produces Be and B by spallation
reactions between CNO nuclei and $p$ and $\alpha$ particles.  If this CCR
nucleosynthesis scenario of $^6$Li production leads to overproduction
of Be and B nuclides against observations in metal-poor stars, it cannot
be achieved in the real universe.  This study is devoted to checking if the CCR
nucleosynthesis as a mechanism of $^6$Li production is consistent with observations of Be and B in
metal-poor stars, and finding constraints on conditions of the CCR
nucleosynthesis.

Abundances of Be and B are observed in MPHSs.  A trend of $^9$Be
abundance is found such that Be increases linearly as Fe during the
course of Galactic evolution~\citep{boe1999}.  The first report on
observation of beryllium in two very metal-poor stars with the
Ultraviolet and Visible Echelle Spectrograph (UVES) mounted on the ESO
VLT Kueyen telescope~\citep{pri2000a} said that the
trend of beryllium with metallicity keeps decreasing at lower
metallicities with no evidence of flattening.  \citet{pri2000b} found that the very metal deficient star G 64-12 ([Fe/H]=$-3.3$) has Be of
log (Be/H)$=-13.10~\pm$~0.15 dex, which is significantly higher than
expected from the previous trend, and claimed that this high [Be/Fe] ratio
may suggest a flattening in the beryllium evolutionary trend at the
lowest metallicity end or the presence of dispersion at early epochs of
the Galactic evolution.  On the other hand,~\citet{boe2006} found that G64-37 with [Fe/H]=$-3.2$ has a Be abundance which is
consistent with the Be-Fe trend, and suggested that different Be values
are indicative of a Be dispersion even at the lowest metallicities.

B abundances in metal-poor stars have been estimated by observations
with the Goddard High Resolution Spectrograph (GHRS) on board the {\it Hubble
Space Telescope} ({\rm HST})~\citep{dun1997,gar1998,pri1999,cun2000}.  The boron abundances are also
found to show a linear increase with a slope of $\sim$~1 with respect to
metallicity, and there is no signature of a primordial plateau abundance.

These trends of Be and B abundances are explained by Galactic CR nucleosynthesis models of different types.  One is the acceleration of
metal-rich CRs, probably freshly synthesized matter at SNe followed by the primary reactions between CR accelerated CNO nuclides and
interstellar nuclides $p$ and $\alpha$,
i.e. [CNO]$_{\rm CR}$+[$p\alpha$]$_{\rm ISM}
\rightarrow$[LiBeB]~\citep[e.g.][]{ram1997}.  This mechanism leads to the same-rate increase of BeB and
O.  \citet{fie2000} have suggested that the secondary reactions between CR accelerated $p\alpha$ and
interstellar CNO i.e. [$p\alpha$]$_{\rm CR}$+[CNO]$_{\rm ISM}
\rightarrow$[LiBeB], perhaps without any contribution of the primary
reactions, would explain the BeB to Fe trend if the ratio O/Fe increases
toward low metallicity.  \citet{val2002} have shown that the
multi-zone (halo, thick disk, and thin disk) Galactic evolution model
including only the secondary reactions can reproduce the linear trend
without fine-tuning. 

In this work, I adopted Model 1 and the rapid burst model of \citet{dai2006} for
the cosmic chemical evolution model.  This chemical evolution model
and other inputs, as well as calculation of light element evolution
are explained in Sec. 2.  I present results of the CCR nucleosynthesis in
Sec. 3, and discuss this study in Sec. 4.  I summarize the CCR
production of LiBeB in Sec. 5.

\section{MODEL}

\subsection{Cosmic SN Rate}

I adopt the cosmic SFRs in a chemical evolution model
given by \citet{dai2006}.  I take Model 1 and the rapid burst
model, which include formation of stars with masses between 40 and 100
$M_{\sun}$ in the early phase of the universe.  As their best model, a parameter, the minimum mass $M_{\rm min}$ of dark matter halos of star-forming
structures is determined to be $10^7~M_{\sun}$ for chemical evolution of
structures. The adopted models are the same as~\citet{rol2006} use.  The
birthrate function is given by
\begin{equation}
B(m,t,Z)=\phi_1(m)\psi_1(t)+\phi_2(m)\psi_2(Z),
\label{eq1}
\end{equation}
where $m, t, Z$ are the mass of star, the age of the universe, and the
metallicity, $\phi_1$ and $\phi_2$ are initial mass functions (IMFs) of
the normal and massive-only component of stars, respectively, $\psi_1$
and $\psi_2$ are SFRs for respective components.
The normal component is given such that its mass range is from 0.1 to
100 $M_{\sun}$.  The massive component is active only at high redshift,
and its mass range is from 40 to 100 $M_{\sun}$.  The IMF of both modes
is given by a power law of mass with an index like the Salpeter's,
\begin{equation}
\phi_i(m)\propto m^{-(1+x)},
\label{eq2}
\end{equation}
with $x=1.3$.  The amplitudes of the IMFs are normalized respectively as
\begin{equation}
\int_{m_{\rm inf}} ^{m_{\rm sup}} dm m \phi_i(m)=1,
\label{eq3}
\end{equation}
where $m_{\rm inf}$ and $m_{\rm sup}$ are the lower and upper ends for
the mass range of each mode.

The normal-mode SFR is given by
\begin{equation}
\psi_1(t)=\nu_1 M_{\rm struct} \exp(-t/\tau_1),
\label{eq4}
\end{equation}
where $\nu_1=0.2$~Gyr~$^{-1}$ describes the efficiency of the star
formation, and $M_{\rm struct}$ and $\tau_1=2.8$~Gyr are mass of the
structure and timescale, respectively. 

In Model 1, the massive-mode SFR is given by
\begin{equation}
\psi_2(t)=\nu_2 M_{\rm ISM} \exp(-Z_{\rm IGM}/Z_{\rm crit}),
\label{eq5}
\end{equation}
where $\nu_2=80$~Gyr~$^{-1}$ is related to the efficiency of star
formation.  $M_{\rm ISM}$ is the baryonic mass of the ISM.  $Z_{\rm IGM}$ is the metallicity of the medium between
the collapsed structures (identified as the IGM), and
$Z_{\rm crit}=10^{-4} Z_\sun$~determines the effective epoch of the end
of Pop III star formation.  In contrast, in the rapid burst
model,the massive mode star formation occurs as an instantaneous event
at redshift $z=16$.  Since the SFR in the model has an instantaneous
bump and I found difficulty reading it from Fig. 13
of~\citet{dai2006}, I fixed the SFR so that it is $\sim 20 M_\sun {\rm
yr}^{-1} {\rm Mpc}^{-3}$ during $3\times 10^6$~yr~\citep{rol2006}.

\subsection{LiBeB Production in the Homogeneous Universe}

I calculate abundances of light elements (LiBeB) produced in the
homogeneous universe through the interaction between fast nuclei
accelerated in the SN shocks and background nuclei.  The picture of
CCRs, i.e., the total kinetic energy of CRs and the propagation is identical to
that of~\citet{rol2006}.

\subsubsection{CCR Energy and Its Spectrum}

The total kinetic energy given to accelerated CRs by SN explosions is 
\begin{eqnarray}
{\cal E}_{\rm SN}(z)&=& (1+z)^3 \nonumber \\
&\times&\int_{{\rm max}(8M_{\sun},m_{{\rm d}}(t))}^{m_{\rm sup}} dm  \sum_{i=1}^2 \phi_i(m) \psi_i(t-\tau(m)) {\cal E}_{\rm CR}(m),\nonumber \\
\label{eq6}
\end{eqnarray}
where $m_{\rm d}(t)$ is the mass of stars with lifetime $t$, $\tau(m)$
is the lifetime of a star of mass $m$, and ${\cal E}_{\rm CR}(m)$ is the
energy imparted to CRs per SN with its initial mass
$m$.  \citet{dai2004, dai2006} adopt stellar lifetimes $\tau(m)$ from
\citet{mae1989} for intermediate-mass stars ($<8~M_\sun$), and
those from \citet{sch2002} for massive stars ($8~M_\sun<m<100~M_\sun$).

\citet{rol2006} give ${\cal E}_{\rm CR}(m)$ after some
assumptions to calculate the total kinetic energy ${\cal E}_{\rm
SN}(z)$.  Core collapse SNe are supplied with energy by
the gravitational collapse of cores.  Almost all of the energy generated
from core collapse, $E_{\rm CC}$ is transferred out by neutrinos.  Only
1~\% will be given to the energy of SN explosions.
\citet{rol2006} use a parameterization
\begin{equation}
{\cal E}_{\rm CR}(m)=\frac{\epsilon E_{\rm CC}(m)}{100},
\label{eq7}
\end{equation}
where $\epsilon$ is the fraction of SN explosion energy imparted
to CRs.  Assumptions related to $E_{\rm CC}$ are as follows:  Every star of mass $m>8~M_\sun$ explodes as SN.  Stars of mass
$8~M_\sun < m < 30~M_\sun$ make neutron stars of mass 1.5~$M_\sun$ at
core collapses and $E_{\rm CC}=3\times 10^{53}$~ergs.  Stars of mass
$30~M_\sun < m < 100~M_\sun$ become black holes with masses very similar
to their helium core mass~\citep{heg2003}.  The mass of the helium
core is $M_{\rm He}=13(m-20~M_\sun)$~\citep{heg2002}.  In this case
$E_{\rm CC}$ is proportional to the mass of the black hole, and
$E_{\rm CC}=0.3M_{\rm He}$ is assumed.

I calculate light element production in the formalism of
\citet{mon1977} and \citet{rol2005, rol2006}.  The proper source function of
SN CRs $Q_i(E,z)$ in a rapid burst at redshift $z_s$ (corresponding time
$t_s$) is defined as
\begin{eqnarray}
Q_i(E,z)=(1+z_s)^3 C(z_s)\frac{\phi_i(E,z_s)}{\beta}\delta(t-t_s)\nonumber \\
~~~~~~~~~~~~~~~~~~~~~~~~[({\rm GeV/nucleon})^{-1}~{\rm cm}^{-3} {\rm s}^{-1}],
\label{eq8}
\end{eqnarray}
where $E$ and $\beta$ are the kinetic energy and velocity of CRs,
respectively, and I define the CR injection spectrum of nuclide $i$ as
\begin{equation}
\phi_i(E,z_s)=K_{ip}^{\rm CR}(z_s) \frac{1}{(E(E+2E_0))^{\gamma/2}},
\label{eq9}
\end{equation}
where $K_{ip}^{\rm CR}$ is the ratio of number abundance of $i$ to that
of $p$, i.e. $i/p$ of CRs, and $E_0=938$~GeV is the nuclear mass energy per
nucleon.  The amplitude of the source function is set so that SNe from
both the normal mode (Pop II) and the massive mode (Pop III) stars supply the CR energy,
\begin{equation}
{\cal E}_{\rm SN}(z) =  \int_{E_{\rm min}}^{E_{\rm max}} E\,\sum_i\,Q_i(E,z)\,dE. 
\label{eq10}
\end{equation}
I take $E_{\rm min}=0.01$~MeV, $E_{\rm max}=10^6$~GeV, and CR spectral
index $\gamma=3$ as \citet{rol2006} do.  Since the evolution of CR
confinement by a magnetic field is difficult to estimate~\citep{rol2006},
as a first step, I assume that the CR confinement is ineffective in the
early universe, so that all CRs generated by SNe in structures
immediately escape from structures to the IGM.  In this case, there is
uniformity of the CR density in the universe.

\subsubsection{Primary Light Element Production by SN CRs}

I define the number density of a CR species $i$ of energy $E$ at
redshift $z$ as $N_i(E,z)$ [in cm$^{-3}$ (GeV/nucleon)$^{-1}$].  In
order to delete the volume changing effect by cosmic expansion, I define
the relative number abundance to that of the background proton $n_{\rm
H}(z)$,
\begin{equation}
N_{i,{\rm H}}(E,z)\equiv N_i(E,z)/n_{\rm H}(z). 
\label{eq11}
\end{equation}
The transport equation for $N_{i,{\rm H}}$, under the isotropic condition, is~\citep{mon1977}
\begin{equation}
\frac{\partial N_{i,{\rm H}}}{\partial t} + \frac{\partial}
{\partial E}(bN_{i,{\rm H}}) + \frac{N_{i,{\rm H}}}{T_{\rm D}} = Q_{i,{\rm H}},
\label{eq12}
\end{equation}
where $b(E,z)\equiv (\partial E/\partial t)$ is the energy loss rate
[(GeV/nucleon)~s$^{-1}$] for cosmic expansion or ionization, and $T_{\rm
D}(E,z)$ is the
lifetime against destruction.  $Q_{i,{\rm H}}(E,z)\equiv Q_i(E,z)/n_{\rm
H}(z)$ is the normalized comoving source function.

The expansion loss and ionization loss are expressed in a product of
energy-dependent term and a redshift-dependent one, $b(E,z)=-B(E)f(z)$.  These
terms are given in \citet{mon1977}. The redshift-dependent term of
the expansion loss is $f_{\rm E}=(1+z)^{-1} |dz/dt|H_0^{-1}$, where
$H_0$ is the Hubble constant.  I assume the standard $\Lambda$CDM model
with its parameters from WMAP three-year data~\citep{spe2007}\footnote{http://lambda.gsfc.nasa.gov} ,
$h=H_0/(100~{\rm km~s}^{-1}~{\rm Mpc}^{-1})=0.704$, $\Omega_b
h^2=0.022$, $\Omega_m =0.27$, $\Omega_\Lambda =0.73$.  The ionization
loss rate is from the fitting formula in \citet{men1971} with
the number fraction of $^4$He, He/H=0.08.  $T_{\rm D}=(n_{\rm H}(z)
\sigma_{{\rm D},i} \beta)^{-1}$ is estimated with the nuclear destruction
cross section $\sigma_{{\rm D},i}$ from \citet{ree1974}.

I define $z^\star(E,E',z)$ as in \citet{mon1977},
\begin{equation}
\frac{\partial z^\star}{\partial E}=-\frac{1}{B(E)f(z)}\left|\frac{dz}{dt}\right|\frac {\partial z^\star}{\partial z}.
\label{eq13}
\end{equation}
A physical interpretation is that a CR particle with energy $E$ at
redshift $z$ had an energy $E'(\geq E)$ at redshift $z^\star(E,E',z)$
before experiencing energy loss.  Thus $z^\star(E,E,z)=z$ is satisfied.
CCR particles with energy $E$ at $z$ originate in those with
$E'_s(E,z,z_s)$ at $z_s$.  $E'_s(E,z,z_s)$ satisfies an equation,
$z^\star(E,E'_s,z)=z_s$.  $z^\star(E,E',z)$ is obtained by integrating
Eq.~(\ref{eq13}) applying the greater loss process to $b$
assuming that the process with the greater rate of $b(E,z)$ is dominant all
the way from redshift $z^\star$ to $z$.

The transfer equation is solved~\citep{mon1977} to obtain the CCR
energy spectrum from a CR burst at $z_s$,
\begin{eqnarray}
\hspace{-10pt}\Phi_{i, {\rm H}}(E,z,z_s) &=& C(z_s) \frac{\phi_i(E'_s,z_s)}{n_{\rm H}^0}\frac{\beta}{\beta'} \left|\frac{d{z}}{d{t}}\right|_{z_s}\nonumber \\
&&\times\frac{\exp{(-\xi(E,E'_s,z))}}{|b(E,z_s)|}\,\frac{1}{\left|\partial z^\star/\partial E'\right|_{{E'_s}}},
\label{eq14}
\end{eqnarray}
where $\Phi_{i,{\rm H}}(E,z,z_s)\equiv \Phi_i(E,z,z_s)/n_{\rm H}(z)$ is
the normalized flux of $i$ per comoving volume with $\Phi_i(E,z,z_s)\equiv
\beta N_i(E,z)_{z_s}$, $\beta$ and $\beta'$ are the velocities corresponding
to energy $E$ and $E'_s$, respectively.  $n_{\rm H}^0$ is the present
average number density of protons in the universe.  $\xi$ is an effect
resulting when the nuclear destruction is considered, and given as
\begin{equation}
\xi(E,E'_s,z)\hspace{-2pt}
=\hspace{-4pt}
\int_E^{E'_s}\hspace{-4pt}
 \frac{dE''}{\left|b(E'',z^\star(E,E'',z)) T_{\rm D}(E'',z^\star(E,E'',z))\right|}.
\label{eq15}
\end{equation}
After analysis with Eq.~(\ref{eq13}), one can estimate $|\partial
z^\star / \partial E'|_{E'=E'_s}=|b(E'_s,z_s)|^{-1} |dz/dt|_{z=z_s}$ and
find an expression for $\Phi_{i, {\rm H}}$
\begin{equation}
\Phi_{i, {\rm H}}(E,z,z_s) = C(z_s) \frac{\phi_i(E'_s,z_s)}{n_{\rm H}^0}\frac{\beta}{\beta'} \frac{|b(E'_s,z_s)|}{|b(E,z_s)|} e^{-\xi(E,E'_s,z)}.
\label{eq16}
\end{equation}

The production rate of light element $l$ of energy $E$, produced at
redshift $z$ is given by
\begin{eqnarray}
\hspace{-4pt}\frac{\partial N_{l,\,{\rm H}}(E,z,z_s)}{\partial t}\hspace{-2pt} &=& \sum_{i,j} \hspace{-3pt}
\int \hspace{-4pt}
\sigma_{ij \rightarrow l}(E,E')n_j(z)\Phi_{i,{\rm H}}(E',z,z_s)\, dE' \nonumber \\
 &=& \sum_{i,j} \hspace{-3pt}
\int \hspace{-4pt}
\sigma_{ij \rightarrow l}(E,E') K_{jp}^{\rm IGM}(z)\Phi_i(E',z,z_s)\, dE',\nonumber \\
\label{eq17}
\end{eqnarray}
where $\sigma_{ij \rightarrow l}(E,E')$ is a cross section of a process between
a CR nuclide $i$ with energy per nucleon $E'$ and a background species $j$ to make a given light
element $l$ with $E$, and $n_j(z)$ and $K_{jp}^{\rm IGM}(z)$ are
background number abundance of a nuclide $j$ and number ratio of $j$ to
proton, respectively.  When the destruction of the light element $l$
after production is neglected, the total production rate is calculated as
\begin{eqnarray}
&&\int \frac{\partial N_{l,\, {\rm H}}(E,z,z_s)}{\partial t} dE \nonumber\\
&&= \sum_{i,j} K_{jp}^{\rm IGM}(z) \int \sigma_{ij \rightarrow l}^{\rm tot}(E') \Phi_i(E',z,z_s)\, dE',
\label{eq18}
\end{eqnarray}
where $\sigma_{ij \rightarrow l}^{\rm tot}(E')$ is the total cross section
of a reaction $i+j \rightarrow l+X$, with any $X$. I adopt cross sections
from \citet{rea1984}, and particularly for the $\alpha+\alpha$ reaction,
exponential-plus-constant cross section for $l=^6$Li and exponential one
for $l=^7$Li from \citet{mer2001}.  The resulting light element
abundance is obtained as the CR production added to the BBN yield.  The
yield by CR nucleosynthesis is the integration of those produced at $z'$
from CRs generated at $z_s$ over $z'$ and $z_s$, thus
\begin{eqnarray}
\left(\frac{l}{\rm H}\right)_{\rm IGM}\hspace{-15pt}(z)&=& \left(\frac{l}{\rm H}\right)_{\rm BBN} \nonumber \\
&&+\int_z^{z_{\rm max}} dz_s \left|\frac{dt}{dz_s}\right| \int _z ^{z_s} dz' \left|\frac{dt}{dz'}\right| \nonumber \\
&&\times \sum_{i,j} K_{jp}^{\rm IGM}(z') \int \sigma_{ij \rightarrow l}^{\rm tot}(E') \Phi_i(E',z',z_s)\, dE'.\nonumber \\
\label{eq19}
\end{eqnarray}

\subsubsection{Secondary Light Element Production by SN CRs}

I also calculate the LiBeB production in the universe by the secondary
process, i.e., [$p\alpha$]$_{\rm CR}$+[CO]$_{\rm
ISM}\rightarrow$[LiBeB]$_{\rm ISM}$.  Since the C and O abundances of
the ISM in structures are about two orders of magnitude higher than those
of the IGM~\citep[see Fig. 11 in][]{dai2006}, the secondary LiBeB
production in the IGM is not important.  I expect that the LiBeB
abundances in the ISM are enhanced by a contribution of the secondary
process.  In fact, the reactions of [$p\alpha$]$_{\rm CR}$+[CO]$_{\rm
ISM}\rightarrow$[LiBeB]$_{\rm ISM}$ make light elements in the ISM and
the mass accretion to the structures from the IGM dilutes the
ISM abundances in the framework of this model involving a hierarchical
structure formation.  Note that from the assumption that the confinement of CRs by
a magnetic field is ineffective, the CRs do not stay in the structures.

The light element abundances
produced by the secondary reactions are then given with a parameter: the
fraction of baryons at redshift $z$ which are in structures
\begin{equation}
f(z) = \frac{\int_{M_{\rm min}}^{\infty}~dM~M~f_{\rm PS}(M,z)}{\rho_{\rm DM}},
\label{eq20}
\end{equation}
where $f_{\rm PS}(M,z)$ is the distribution function of
halos taken
from the \citet{she1999} modification to the Press-Schechter
function~\citep{pre1974} converted into the mass function~\citep{jen2001} by a code provided by
A. Jenkins (2007, private communication).  I assume that the primordial
power spectral slope is $n$=1, the rms amplitude for mass density
fluctuations in a sphere of radius 8 $h^{-1}$ Mpc is $\sigma_8=0.9$, and
the \citet{bon1984} fit to the transfer function for cold dark
matter is used in generating a mass function.  $\rho_{\rm DM}$ is the
comoving dark matter density of the universe.

The light elements made by the secondary process are contained in the
structures that grow gradually.  The abundance of a light element in the
structures is then given by
\begin{eqnarray}
\left(\frac{l}{\rm H}\right)_{\rm ISM}\hspace{-15pt}(z)&=& \left(\frac{l}{\rm H}\right)_{\rm IGM}\hspace{-15pt}(z) \nonumber \\
&&\hspace{-5pt}+ \frac{1}{f(z)}\int_z^{z_{\rm max}} dz_s \left|\frac{dt}{dz_s}\right| \int _z ^{z_s} dz' \left|\frac{dt}{dz'}\right| \nonumber \\
&&\hspace{-5pt}\times \sum_{i,j} K_{jp}^{\rm ISM}(z') f(z') \int \sigma_{ij \rightarrow l}^{\rm tot}(E') \Phi_i(E',z',z_s)\, dE'.\nonumber \\
\label{eq21}
\end{eqnarray}

\section{RESULTS}

I calculate the light element production in the uniform universe by CCRs
(i.e. neglecting an inhomogeneity of CCRs).  I consider only processes
between the accelerated CRs with the abundance patterns of the structures
in the~\citet{dai2006} model and the background IGM abundance which I assumed
to be of the primordial abundance.  I set the primordial helium abundance He/H=0.08.

\begin{figure}
\begin{center}
\includegraphics[angle=-90,scale=0.5]{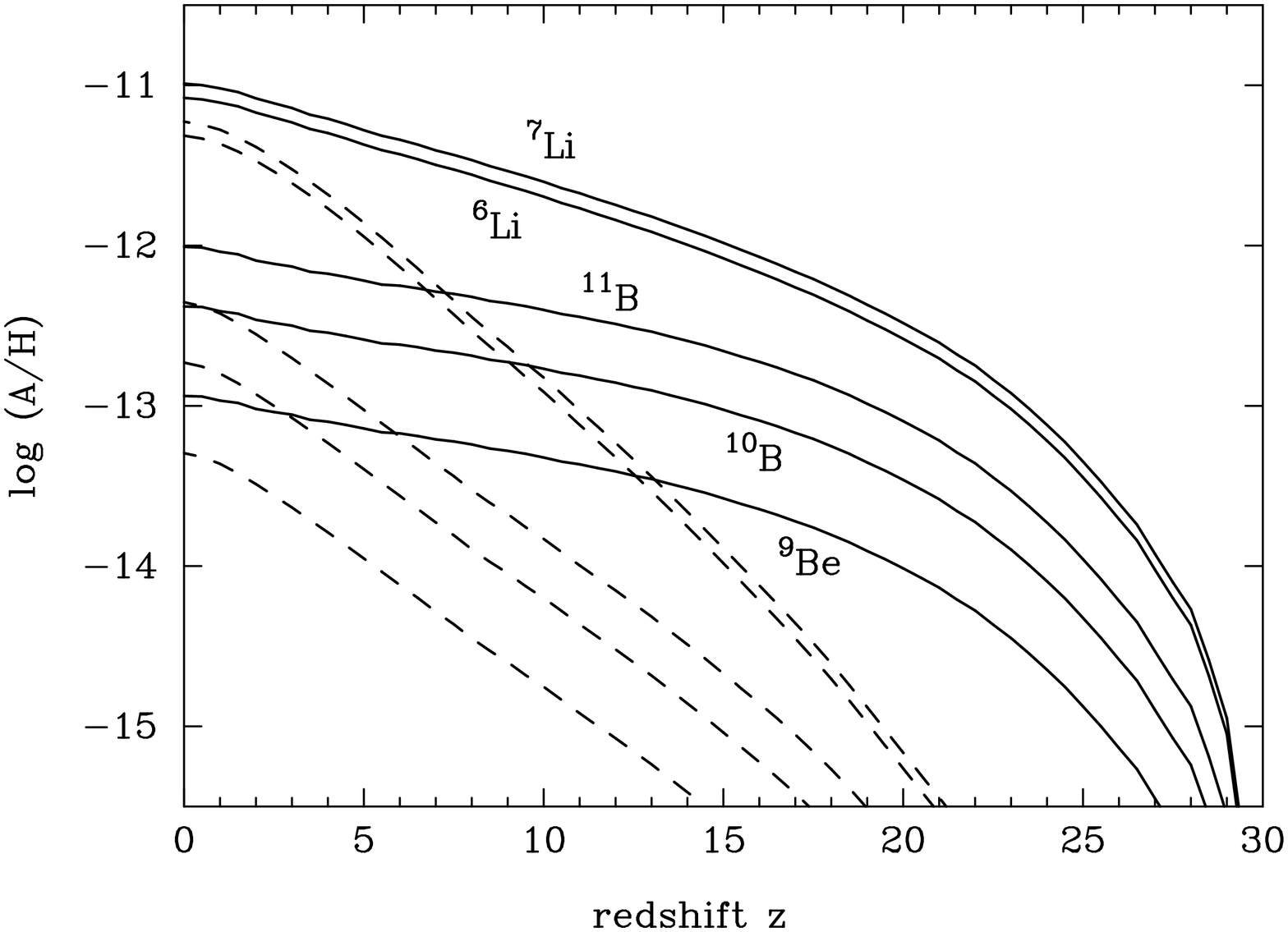}
\end{center}
\caption{Abundances of light elements in the IGM as a function of redshift
 in Model 1 (solid lines). $\epsilon=0.31$ is assumed to result in
 $^6$Li/H=$6\times 10^{-12}$ at $z=3$.  The contribution of the normal
 mode stars only to the light element production is shown by the dashed
 lines.\label{fig1}}
\end{figure}

\begin{figure}
\begin{center}
\includegraphics[angle=-90,scale=0.5]{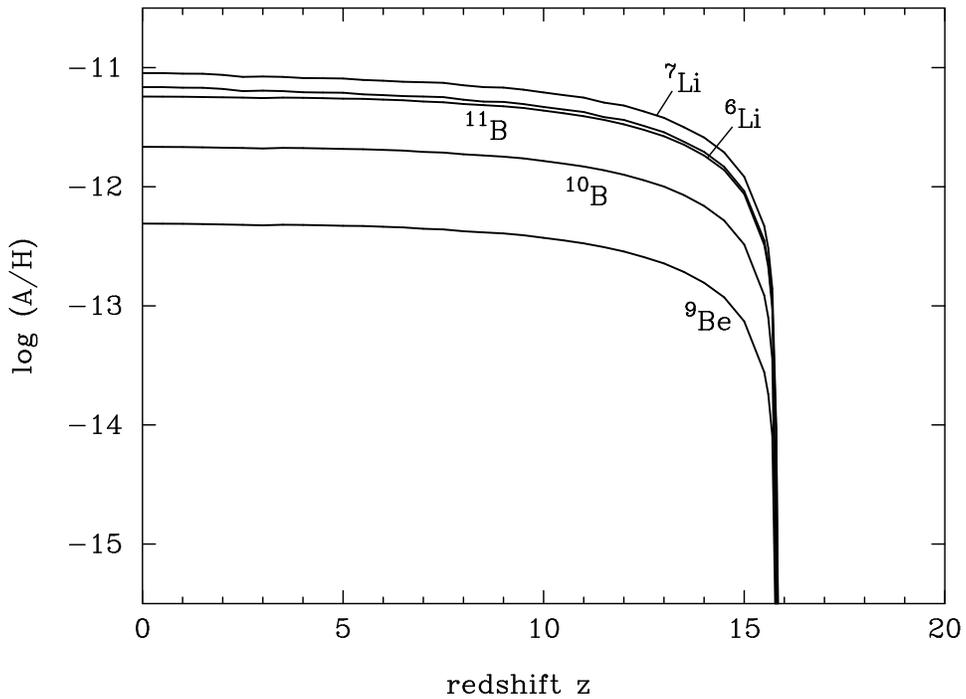}
\end{center}
\caption{Abundances of light elements in the IGM as a function of redshift
 in the rapid burst model. $\epsilon=0.029$ is assumed to result in
 $^6$Li/H=$6\times 10^{-12}$ at $z=3$.\label{fig2}}
\end{figure}

I show the result of light element evolution in Model 1 in Fig.\
\ref{fig1}.  Solid lines correspond to the case where both the normal
and massive modes are included, and dashed lines correspond to the case
in which only the normal mode is included as the energy source.  When the SN
energy is totally given to the CR acceleration ($\epsilon=1$),
$^6$Li/H=$2.0\times 10^{-11}$ at $z=3$ is derived.  In
Fig.\ \ref{fig1}, $\epsilon=0.31$ is assumed so that $^6$Li/H=$6\times
10^{-12}$ at $z=3$ results that is the observed abundance level in
MPHSs.  In the case that the CRs are energized by only the normal mode star
formation, the energy fraction $\epsilon=0.73$ is needed to realize the
observed $^6$Li abundance.  The time evolution of light element
abundances in the rapid burst model is shown in Fig.\ \ref{fig2}.  The
energy fraction given to the CR acceleration is assumed to be
$\epsilon=0.029$, and the abundance of $^6$Li gets $^6$Li/H=$6\times
10^{-12}$ at $z=3$.  These results are very similar to that
in~\citet{rol2006} (See their Fig. 2).  In Model 1, $^6$Li is produced
gradually with decreasing redshift, while in rapid burst model, it is
immediately produced by most part at the burst of star formation.  In my
calculation, the $^6$Li production at high redshift tends to be slightly
more efficient than in~\citet{rol2006}.  This difference might be caused
by different numerical calculation of transport equation for nuclides.

\placefigure{fig1}
\placefigure{fig2}

\begin{figure}
\begin{center}
\includegraphics[angle=-90,scale=0.5]{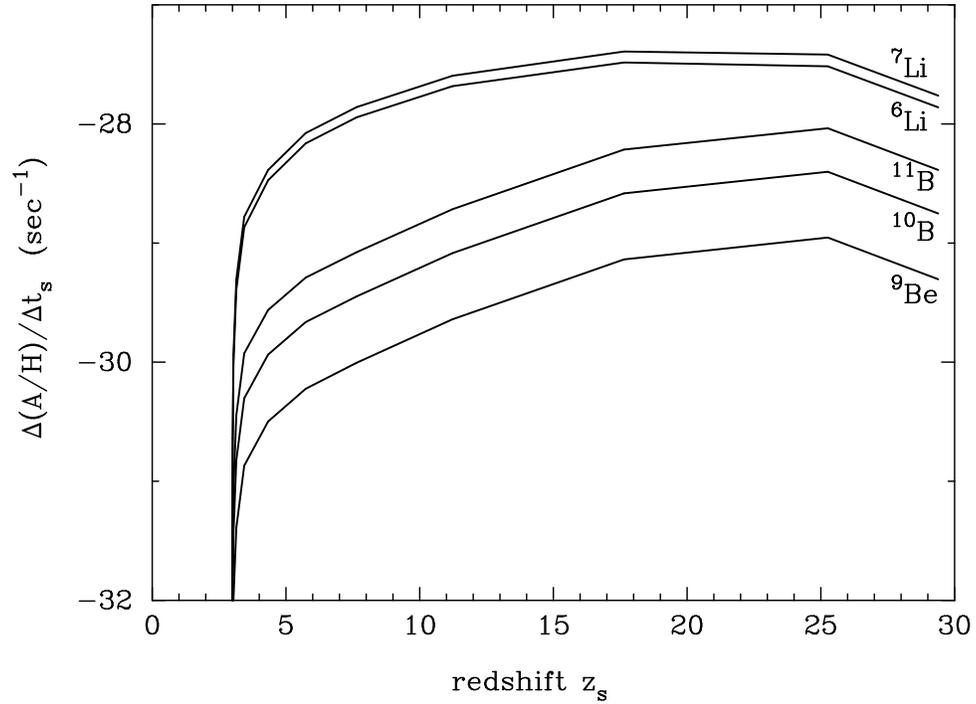}
\end{center}
\caption{Yields of light elements at $z=3$ by CCRs generated at $z_s$ per second
 in Model 1.\label{fig3}}
\end{figure}

Figure\ \ref{fig3} shows the yields by CCRs generated at $z_s$ per second
in Model 1, i.e. $\Delta$($l$/H)/$\Delta t_s$.  The calculated
results show that $^6$Li and $^7$Li are produced mainly by the $\alpha+\alpha$ fusion
reaction, and Be and B production has the strongest contribution from
the O+$p$ (and C+$p$) spallation processes.  I have taken the helium
abundance He/H=0.08, and the O abundance in the ISM of structures in
Model 1 is roughly constant from $z\sim 0$ to $z\sim 20$~\citep[see Fig. 11 of][]{dai2006}.  The injected energy density in CRs is smoothly
increasing as a function of redshift $z$~\citep[Fig. 2 of][]{rol2006}.  This figure therefore reflects the abundance of seed
nuclides and injected energy density.  The CRs generated at $z_s\sim 3$
have little time to react with background nuclear species, so that the
yields go to zero, and a decrease in all yields at $z_s\lesssim 25$
reflects the shape of CR energy density~\citep[Fig. 2 of][]{rol2006}.

\placefigure{fig3}

I assume that MPHSs formed at redshift $z\sim 3$ and that they include
LiBeB elements at the level of the IGM abundances at the time.  If one can
neglect the inhomogeneity of CCRs resulting from the local growth of a
magnetic field, numbers of produced elements are proportional to target
particle numbers.  Consequently, resulting light element abundance $l$/H
does not depend on density there.  Although the metallicities like Fe/H
of metal-poor stars reflect how the metal-enhanced and metal-deficient
gas are mixed before star formation, the light element abundances are
constant for the same formation epoch.  In this case, there appear
primordial plateaus on the plot of abundances as a function of
metallicity [Fe/H].   The energy fractions given to the CR acceleration,
$\epsilon=0.31$ for Model 1 and $\epsilon=0.029$ for the rapid burst
model realize the observed abundance of $^6$Li in MPHSs at $z=3$.  These
fractions are reasonable, considering that the energy of Galactic CRs can
be covered by $10-30$ percent of supernova remnant (SNR) energy~\citep{dru1989}.  If
$^6$Li observed in MPHSs is produced mainly by this CCR nucleosynthesis,
these stars include Be and B which had been coproduced.

\begin{figure}
\begin{center}
\includegraphics[angle=-90,scale=0.5]{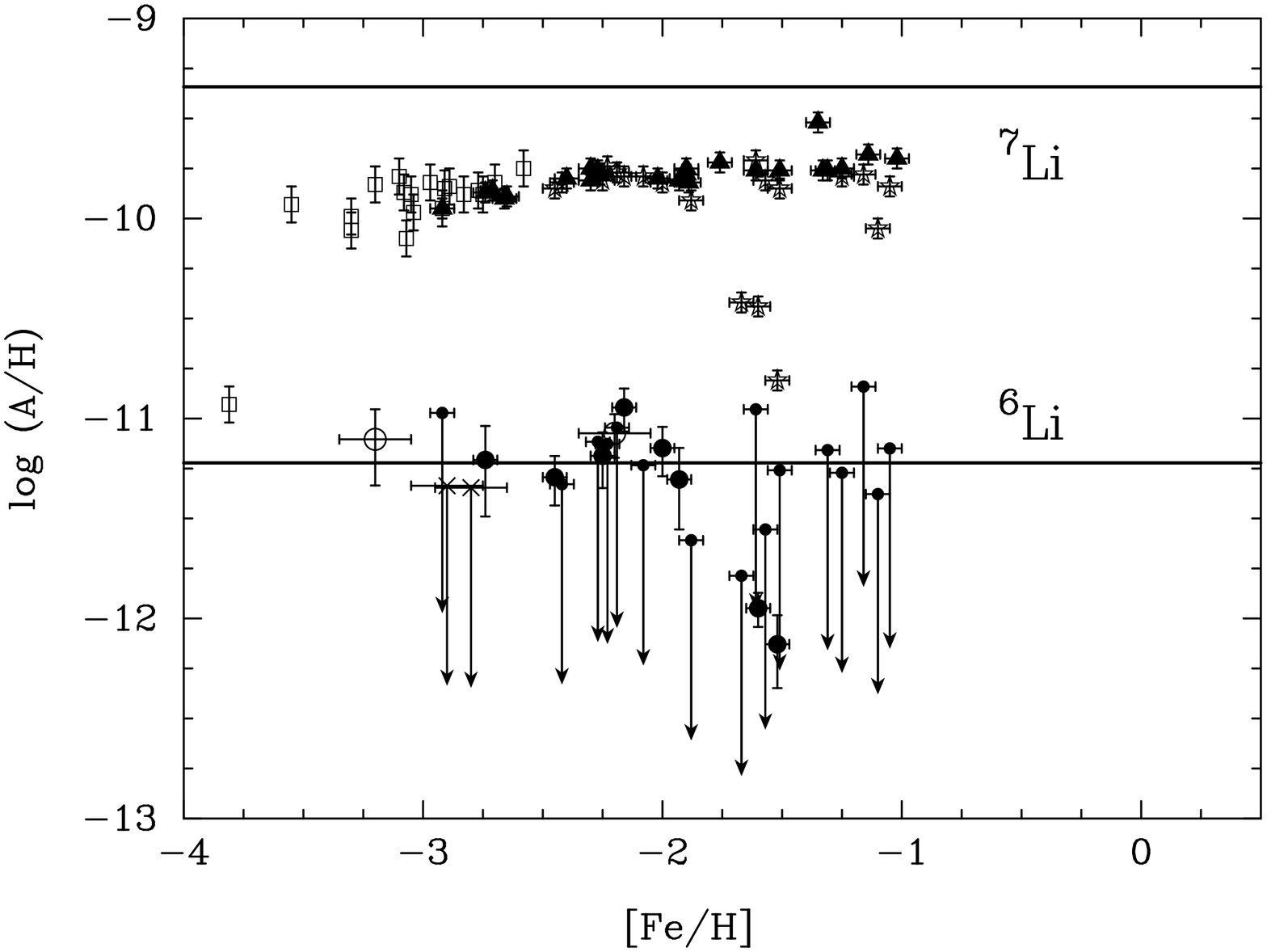}
\end{center}
\caption{Plateau abundances of lithium isotopes produced by the CCR
 nucleosynthesis in Model 1
 with the accelerating efficiency $\epsilon=0.31$.  $^7$Li data are from
 \citet[filled triangles]{asp2006}, \citet[open squares]{bon2007}, and
 \citet[open stars]{shi2007}.  $^6$Li data are from \citet[large filled circles to detections, small filled circles
 to upper limits]{asp2006} and \citet[open circles to detections,
 crosses to upper limits]{ino2005}.\label{fig4}}
\end{figure}

In Fig.\ \ref{fig4} the plateau levels of $^6$Li and $^7$Li in Model 1
are drawn with observational data points in a plot of abundances as a
function of metallicity [Fe/H].  As for the $^7$Li abundance, the BBN
prediction is calculated using the Kawano code~\citep{kawano1992} with the
use of the new world average of the neutron lifetime~\citep{mat2005}.  I take the energy density of baryons in the universe given
by WMAP first year data analysis~\citep{spe2003} that is $\Omega_b
h^2=0.0224\pm 0.0009$.  This value corresponds to the
baryon-to-photon ratio $\eta=6.1\times 10^{-10}$.  The predicted
abundance is then ($^7$Li/H)$_{\rm BBN}$=4.5$\times 10^{-10}$.  The
$^7$Li production by the CCR nucleosynthesis is a minor addition to the
BBN result.  The figure for the rapid burst model is very similar to
Fig.\ \ref{fig4}.  The factor of about three difference between the BBN
prediction and the observation of $^7$Li abundance is apparent.

\placefigure{fig4}

\begin{figure}
\begin{center}
\includegraphics[angle=-90,scale=0.5]{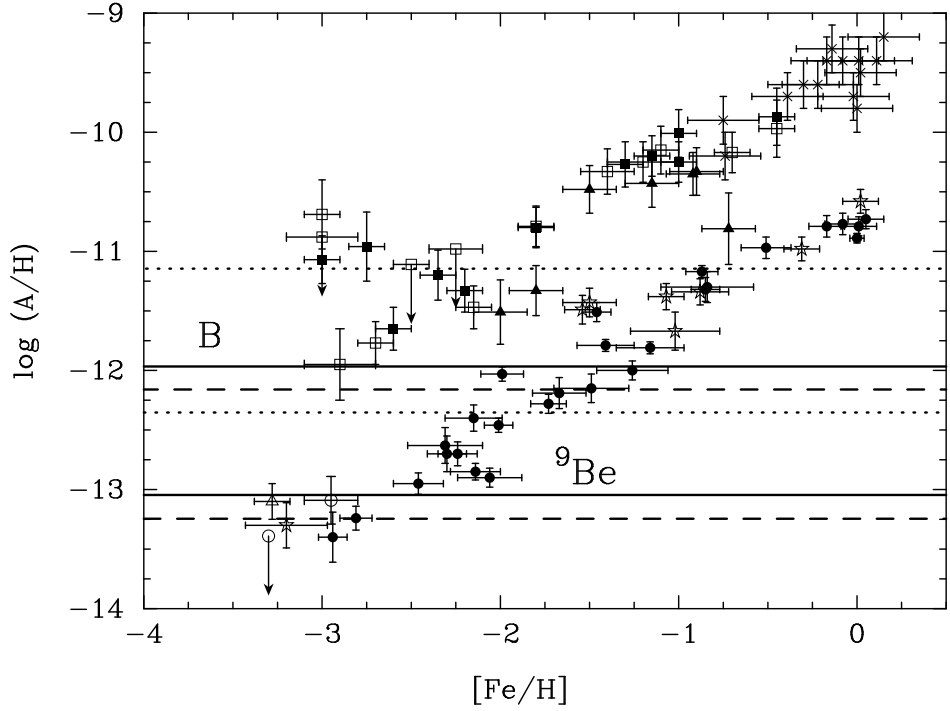}
\end{center}
\caption{Plateau abundances of Be and B produced by the CCR nucleosynthesis
 in Model 1 (solid lines) with the accelerating efficiency
 $\epsilon=0.31$.  The case that the normal mode star formation alone is
 considered as the CR energy source corresponds to the dashed lines with
 $\epsilon=0.73$ to realize the MPHS value of $^6$Li at $z=3$.  Plateau abundances in the rapid burst model with the accelerating efficiency
 $\epsilon=0.029$ are also shown as the dotted lines.  $^9$Be data are
 from \citet[filled circles]{boe1999}, \citet[open circles]{pri2000a},
 \citet[open triangle]{pri2000b}, and \citet[open stars]{boe2006}.  B
 data are from \citet[filled squares]{dun1997}, \citet[open
 squares]{gar1998}, \citet[filled triangles]{pri1999}, and
 \citet[crosses]{cun2000}.\label{fig5}}
\end{figure}

The coproduced abundance levels of Be and B in Model 1 are shown in Fig.\ \ref{fig5}
with observational data points.  It is interesting that the predicted
abundance levels of Be and B are located at nearly the lowest point ever
detected.  This model predicts primordial plateau abundances of Be and B
as an analog of the likely $^6$Li plateau.  If future observation of
beryllium in metal-poor stars catches evidence of a plateau abundance,
it would be explained by this CCRs origin.  When the normal mode star
formation alone is taken as the CR energy source, the result is consistent
with the observed abundance of $^6$Li in MPHSs at $z=3$ with
$\epsilon=0.73$.  This case corresponds to dashed lines in Fig.\
\ref{fig5}.  In the same way, the coproduced abundances in the rapid
burst model with $\epsilon=0.029$ are shown as the dotted lines.  This calculation shows that Be and B abundances are somewhat
higher in the rapid burst model than in Model 1 relative to that of
$^6$Li.  This calculation seems possibly to contain an error of the
order of $\sim 10$~\% in Be and B abundances associated with the
reading of a very brief burst of SFR~\citep[Fig. 13 of][]{dai2006} and
resulting sudden enrichment of C and O in the ISM of
structures~\citep[Fig. 16 of][arXiv:astro-ph/0509183]{dai2006} as input
profiles in this calculation.  In the cosmic chemical evolution model
of~\citet{dai2006}, the rapid burst of star formation occurs in a very
brief time and C and O abundances in structures increase accordingly
reflecting the ejection of C and O nuclei by SNe.  Since Be and B
production is sensitive to the CR energy density times C and O
abundances in structures, the fine-meshed time evolution of C and O
abundances as well as CR energy density (which is associated with SN
rate) is necessary to perform a precise calculation.  However, since the rapid burst model gives sudden rises of C and O
abundances, and C and O particles are accelerated before the ISM
abundances can be somewhat diluted by accretion of the IGM, abundances of C and O
in CCRs are relatively high.  Then it is as expected that the rapid
burst model produces more Be and B relative to $^6$Li than in Model 1,
considering that Be and B are produced by spallation processes of C and
O while lithium isotopes are produced mainly by $\alpha+\alpha$ fusion.
I show the results of calculations of light element production in
the CCR nucleosynthesis model in Table \ref{tab1}.  In the second
column, values of energy fraction of SNRs given to CR acceleration
required to produce $^6$Li at the observed level in MPHSs at $z=3$ are
shown.  The resulting abundances of light elements with the CR
acceleration efficiencies in the second column are listed in the third
to seventh columns.

\begin{figure}
\begin{center}
\includegraphics[angle=-90,scale=0.5]{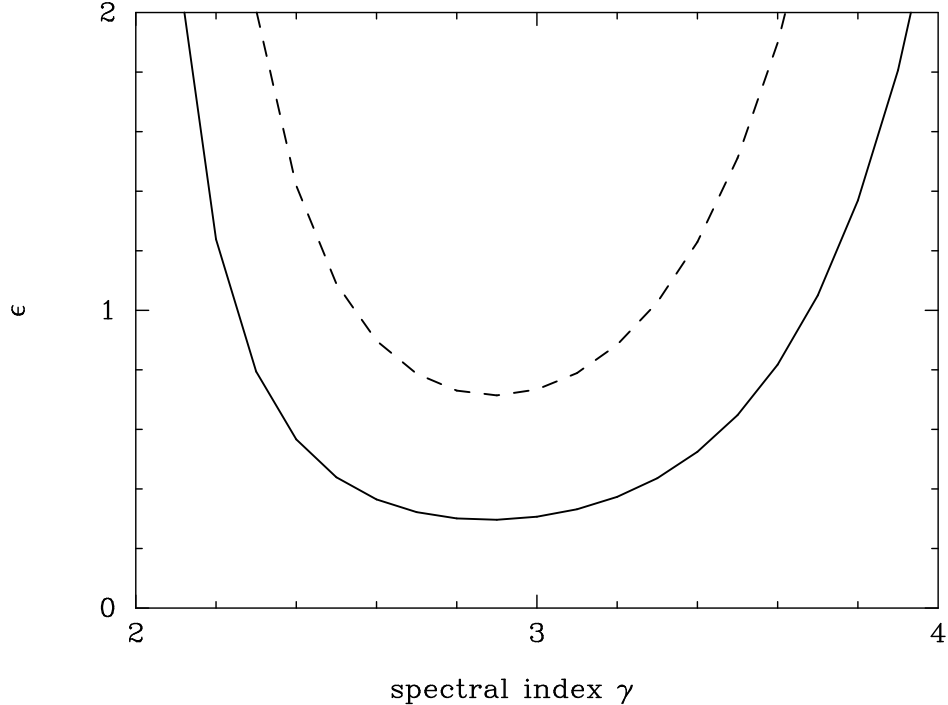}
\end{center}
\caption{Energy fraction of SNRs to CRs, $\epsilon$, as a function of the
 index of the CR injection spectrum, $\gamma$, required to produce
 $^6$Li at the MPHSs level $^6$Li/H=$6\times 10^{-12}$ at $z=3$. The
 solid line corresponds to Model 1 and the dashed line to the case where
 only the normal mode stars are considered as the energy source of
 CCRs.\label{fig6}}
\end{figure}

\placefigure{fig5}
\placefigure{fig6}
\placetable{tab1}

In the calculations so far, the spectral index of the CR injection
spectrum has been fixed to $\gamma=3$.  I show the energy fraction of
SNRs to CRs, $\epsilon$, required to produce $^6$Li at the MPHSs level at
$z=3$ as a function of $\gamma$ in Fig.\ \ref{fig6}. The solid line
corresponds to Model 1 and the dashed line to the case where only the
normal mode stars are considered as the energy source of CCRs. One can
see that the required energy fraction $\epsilon$ is reasonable if
$2.7\lesssim \gamma \lesssim 3.1$.  Meanwhile, smaller spectral index of
$\gamma \lesssim 2.7$ could not produce enough $^6$Li in this model.

\begin{figure}
\begin{center}
\includegraphics[angle=-90,scale=0.5]{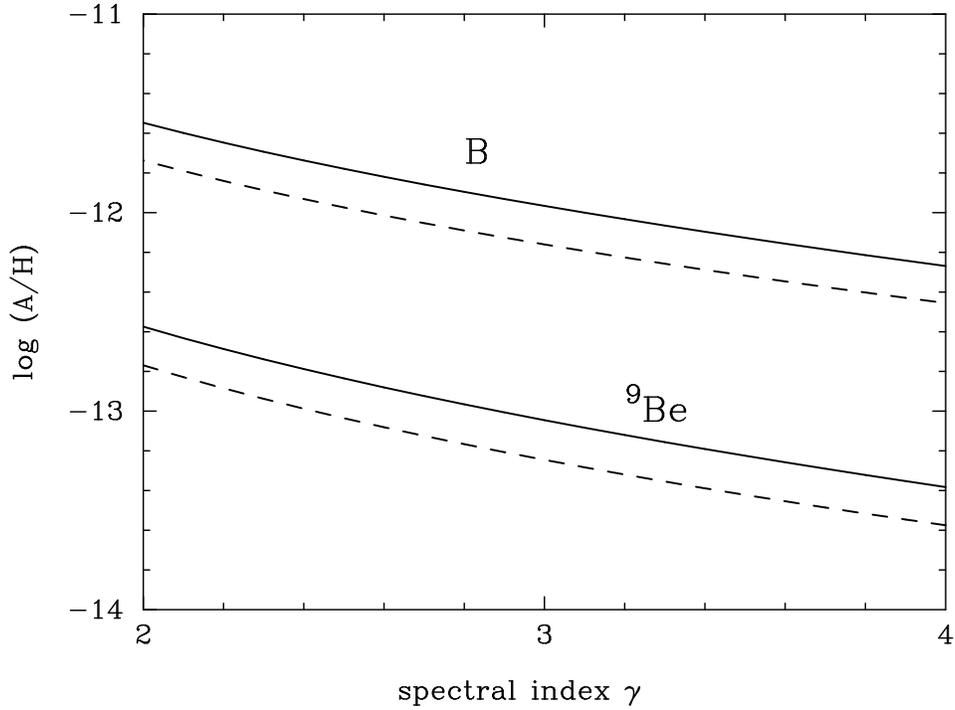}
\end{center}
\caption{Abundances of Be and B at $z$=3 in Model 1 with the CR energy
 fraction $\epsilon$ in Fig.\ \ref{fig6}, when $^6$Li is produced at the
 MPHSs level.  The solid lines correspond to Model 1 and the dashed lines
 to the case where only the normal mode stars are considered as the energy
 source of CCRs.\label{fig7}}
\end{figure}

\placefigure{fig7}

Figure\ \ref{fig7} shows abundances of Be and B at $z=3 $ as a
function of $\gamma$ in Model 1 with the value of $\epsilon$
in Fig.\ \ref{fig6}, i.e., when $^6$Li is produced at the MPHSs level.  It is
found that relative produced abundances of Be and B are decreasing
functions of $\gamma$.  In other words, the steeper the momentum spectrum
of CRs is, the more $^6$Li is produced relatively.  In the region
$2.7\lesssim \gamma \lesssim 3.1$, where the $^6$Li abundance produced
in this model can attain the MPHSs level with a reasonable partition
of SNR energy to CR acceleration, log(Be/H)$\sim -13.1$~-~$-12.9$ and
log(B/H)$\sim -12.0$~-~$-11.8$ are obtained.  These abundances are again
near the least abundances detected in metal-poor stars.

\begin{figure}
\begin{center}
\includegraphics[angle=-90,scale=0.5]{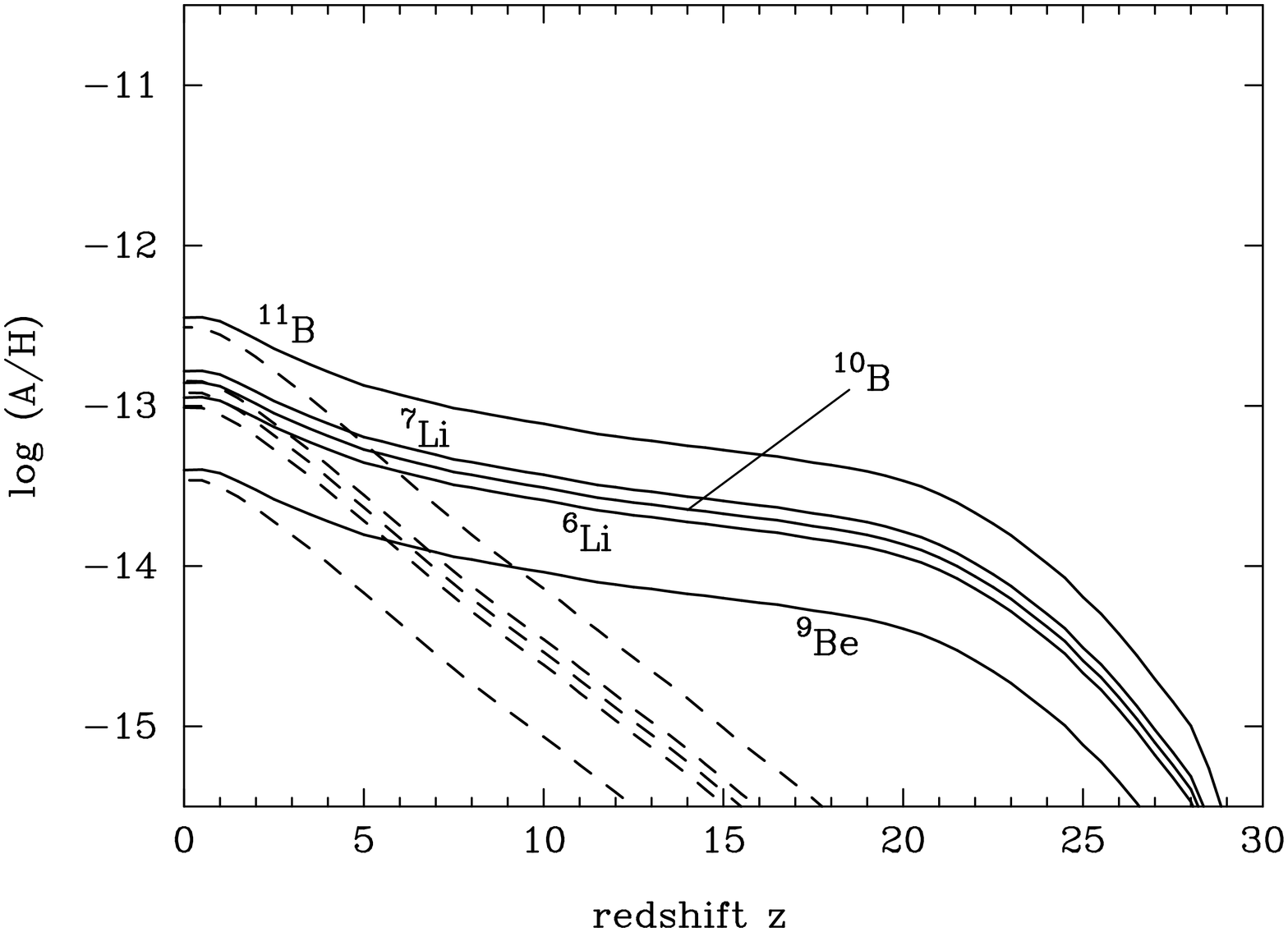}
\end{center}
\caption{Abundances of light elements produced by the secondary process
 in the ISM as a function of redshift in Model 1 (solid
 lines). $\epsilon=0.31$ is assumed to result in $^6$Li/H=$6\times 10^{-12}$ in the IGM at $z=3$.  The contribution of the normal mode stars only to the light element production is shown by the dashed lines.\label{fig8}}
\end{figure}

\placefigure{fig8}

Figure\ \ref{fig8} shows the abundances of light elements produced by the
secondary process in the ISM as a function of redshift in Model 1
calculated by the second term in the rhs of Eq.\
(\ref{eq21}). $\epsilon=0.31$ is assumed to result in $^6$Li/H=$6\times
10^{-12}$ in the IGM at $z=3$.  Solid lines correspond to the case where
both the normal and massive modes are included, and dashed lines
correspond to a contribution of the normal mode as the energy source.
The contributions of primary and secondary processes to the light element
synthesis in the ISM are estimated by the comparison between Fig.\
\ref{fig1} and Fig.\ \ref{fig8}.  Since $^6$Li and $^7$Li are produced mainly
by the $\alpha+\alpha$ fusion which I contained in the calculation of
Fig.\ \ref{fig1}, the contributions of the secondary process are relatively
small.  On the other hand, Be and B have some contribution from the
secondary process.  For example, the Be and B abundances produced by the
secondary process are about 25~\% (65~\%) of those by the primary
process at $z=3$ in Model 1 with both the normal and massive modes
(with only the normal mode).  It is found that the BeB abundance levels
produced by the CCR nucleosynthesis are roughly the same in the ISM and
the IGM.  This contribution from the secondary
process is small in the rapid burst model (about 2.5~\% for Be and B) at
$z=3$.  The fraction of baryons included in the ISM is smaller
at higher redshift, when the energy injection in the rapid burst model
occurs.  It is then easily understood that the light element abundances
produced by the secondary reactions are smaller in the rapid burst model
considering the dilution of light element abundances due to mass
accretion by structure formation in the model.

\section{DISCUSSION}
The resulting abundances from Eq.\ (\ref{eq18}) have a clear dependence on
input quantities as
\begin{equation}
\int \frac{\partial N_{l,\, {\rm H}}(E,z,z_s)}{\partial t} dE \propto \sum_{i,j} {\cal E}_{\rm SN}(z_s) K_{ip}^{\rm CR}(z_s) K_{jp}^{\rm IGM}(z).
\label{eq22}
\end{equation}
I adopted the results of the two models of \citet{dai2006} as the total
CRs energy by SN explosions ${\cal E}_{\rm SN}(z_s)$ and the ratio of
number abundance of $i$ to that of $p$, i.e. $i/p$ of CRs $K_{ip}^{\rm CR}(z_s)$,
which I assume to be the same as the ISM abundance of structures.
These two quantities are uncertain, in fact.  I use the primordial
abundance for the background number ratio of $j$ to $p$ $K_{jp}^{\rm
IGM}(z)$.  This quantity is reasonable and would not contain large
uncertainty.  A large difference in $K_{ip}^{\rm CR}(z_s)$ ($i$=C,O) leads to a
difference in produced Be and B abundances linearly, while $^6$Li and
$^7$Li abundances have little influence since they are mainly produced by
the $\alpha+\alpha$ fusion process.

\citet{suz2001} have developed a model for the evolution of
light elements in the Galaxy, which includes the SN-induced
chemical evolution with contributions from SNe and CRs nucleosynthesis
self-consistently.  They have explained the light element abundances
observed in metal-poor stars using their model with the CR abundances
including SN ejecta 3.5~\% in mass fraction.   The linear relation
between [BeB/H] and [Fe/H] has been obtained by the primary process to
make Be and B.  Consequently their CR abundance includes the C and O in
mass fraction of $(5-7)\times 10^{-3}$ originating from the SN ejecta
throughout the Galactic chemical evolution.  On the other hand, the mass
fraction of C+O is $(3-10)\times 10^{-3}$ in Model 1 of \citet{dai2006} I adopt, which is roughly the same level of abundance as
that \citet{suz2001} used within a factor of $\lesssim 2$.  The
CR abundance I give in this study, therefore, would be appropriate
within a factor of $\sim 2$, even if the real C and O abundances in the
ISM producing CRs were lower than I give, supposing that the CRs have
contribution from the SN ejecta by the fraction inferred
by~\citet{suz2001}.

I assume that all CRs escape from structures to the IGM, and do not
consider a nonuniformity of the CR density in the universe.  As
structures grow in the universe, a magnetic field grows accordingly, and
the CR flux might get inhomogeneous, especially at low redshift, while the overproduction of $^6$Li provides a constraint on the
confinement of CRs in the ISM~\citep{rol2006}.  Further study
including the space distribution and time evolution of a magnetic field
is desirable to estimate the light element abundances produced by the CCR
nucleosynthesis.

The calculated result would also contain an uncertainty from two-step
reactions, i.e., production of nuclide $l_2$ by sequential non-thermal
nuclear reactions: $i+j\rightarrow l_1$ and $l_1+j_2\rightarrow l_2$ with
any nuclides $l_1$ and $j_2$~\citep{ram1997,kne2003}.  \citet{kne2003}
have found that two-step reaction rates are only of the order of 1/10
smaller than one-step reaction rates.  The effect of two-step reactions
in the framework of this study is roughly estimated as follows.  Nuclei produced by nuclear reactions
experience an energy loss and nuclear destruction.  If one take nuclei
with kinetic energy of $E\gtrsim 10$~MeV/nucleon, the expansion loss is
dominant loss process in the considered redshift range.  The time scale
for expansion loss is given by
\begin{eqnarray}
T_{\rm loss}&\sim& \frac{dE}{b_{\rm exp}(E,z)}\nonumber \\
&=&\frac{E+E_0}{H_0 (E+2E_0)}\left[\Omega_m(1+z)^3 + (1-\Omega_m)\right]^{-1/2} \nonumber \\
&=&4.4\times 10^{17}~{\rm s} \frac{E+E_0}{(E+2E_0)}\left[\Omega_m(1+z)^3 + (1-\Omega_m)\right]^{-1/2},\nonumber \\
\label{eq23}
\end{eqnarray}
where $b_{\rm exp}(E,z)$ is the energy loss rate for the cosmic
expansion.  On the other hand, using an empirical formula of nuclear
destruction~\citep[Eq.\ 1 in ][]{let1983}, i.e., $\sigma=44.9
A^{0.7}$~mb, the time scale for nuclear destruction is given by
\begin{eqnarray}
T_{\rm D}&=& (n_{\rm H}(z) \sigma \beta)^{-1} \nonumber \\
&=&8.0\times 10^{20}~{\rm s} \left(\frac{A}{10}\right)^{-0.7} (1+z)^{-3} \beta^{-1}.
\label{eq24}
\end{eqnarray}
Ratios of time scales for nuclides with $A\sim 10$ and
$E=10-10^6$~MeV/nucleon are $T_{\rm loss}/T_{\rm D}\sim 10^{-3}-10^{-2}$
at $z=3$, $3\times 10^{-3} - 3\times 10^{-2}$ at $z=10$, and $10^{-2} -
0.2$ at $z=30$, respectively.  Therefore, the fraction at which nuclei
produced by nuclear spallations experience second nuclear reactions
before losing enough energy is expected to be at most of the order of
$\sim 10$~\%, if the energy spectrum of CRs indicates small amount of
high energy CRs.  I check a fraction of high energy CRs in number, which
would be estimated by
\begin{equation}
F(\gamma)\equiv \frac{\int_{\rm 100~MeV}^{E_{\rm max}} Q_i(E',z)dE'}{\int_{\rm 10~MeV}^{E_{\rm max}} Q_i(E',z)dE'}=\frac{P({\rm 100~MeV})}{P({\rm 10~MeV})},
\label{eq25}
\end{equation}
where I defined
\begin{equation}
P(E)\equiv \int_E^{E_{\rm max}} \frac{E'+E_0}{\left[E'(E'+2E_0)\right]^{(\gamma+1)/2}} dE'.
\label{eq26}
\end{equation}
For $\gamma=2, 3, 4$, $F(2)=0.31$, $F(3)=0.095$, and $F(4)=0.029$ are
obtained, respectively.  The fraction of high energy CRs is thus
relatively low for the range of $2\leq \gamma \leq 4$.  As a
result, the effect of two-step reactions is small and would be at most of the order
of $\sim 10$~\%.

\citet{rol2008} also study the CR production of Be and B by CCRs.  Their
conclusion is very similar to that of this study, and a potentially
detectable Be and B is produced by CCR-induced spallation reactions at
the time of the formation of the Galaxy ($z\sim 3$).  However, there are
some differences of assumptions between the two studies, which are
compared here.  I give the CR spectrum of CO nuclides by the same shape
as those of $p$ and $\alpha$ particles, while~\citet{rol2008} give it by
a broken power law which matches the observed present-day Galactic CR
spectrum.  Moreover, I give the abundances of CR by those of structures
in the cosmic chemical evolution model of~\citet{dai2006},
while~\citet{rol2008} give them by those of structures
in~\citet{dai2006} model multiplied by abundance enhancement factors of
present Galactic CR fluxes.  I assume that the CR confinement by a
magnetic field is ineffective in the early universe, and that all CRs
generated by SNe in structures escape to the IGM, while~\citet{rol2008}
apply a shape of CR diffusion coefficient, which leads to some degrees
of the nuclear destruction of CRs in structures and the application of a
broken power law in the CR energy spectrum of CO nuclides.  Their
diffusion coefficient is derived assuming that it is given by that in a
magneto-hydrodynamic (MHD) turbulence, and that the magnetic energy
density is proportional to the thermal energy, and that a characteristic
length scale for the magnetic field is given by the local Jeans scale.
It is interesting that the two studies using different assumptions for
the uncertain physical inputs arrived at the similar conclusions.

I comment on a difference between the results of this CCR production of
the light elements and those of the flare production model~\citep{tat2007}.  The CCR nucleosynthesis leads to the production of Be
and B at the lowest level ever detected, in this model calculation.  The
nucleosynthesis on the main sequence stars triggered by flare-accelerated nuclides, on the other
hand, results in the negligible productions of $^7$Li, Be and B, which
are proportional to the metallicity and exist at very low abundance
levels under the observed linear trend as a function of metallicity.
Therefore if we observe a signature of the primordial origin of Be and B
by measurements of MPHSs, the production mechanism of Be and B would not
be the flare-energized nuclear reactions after the star formations, and
it would be thought that the plateau abundances of Be and B originate
in the CCR production and $^6$Li has been coproduced at pregalactic
phase by the CCRs.

\section{CONCLUSIONS}
The recent observations of MPHSs reveal the probable existence of high
plateau abundance of $^6$Li, which is about a thousand times higher than
predicted in the standard BBN model.  Since the standard Galactic
chemical evolution model with the Galactic CR nucleosynthesis
gives lower values of $^6$Li abundance at the metallicities of the observed MPHSs, some mechanism must have produced $^6$Li existing
in the surface of MPHSs.  As a candidate of a $^6$Li production
mechanism, the early burst of CRs has been proposed~\citep{rol2005}, and the nucleosynthesis by the CRs from SN
explosions is calculated~\citep{rol2006} in a detailed model of
cosmic chemical evolution~\citep{dai2006} which satisfies various observational
constraints including an early reionization of the universe.  \citet{rol2006} have found that the $\alpha+\alpha$
fusion reaction can produce $^6$Li to the level observed in MPHSs.

I calculate the cosmological cosmic ray nucleosynthesis of Be and B
isotopes as well as $^6$Li and $^7$Li with the use of Model 1 and
the rapid burst model in
\citet{dai2006}.  It is assumed that all CRs produced by SNe in the
ISM escape to the IGM and the CR intensity is always homogeneous in the
universe.  I found that when Model 1 (the rapid burst model) of \citet{dai2006} is adopted for the SFR of the universe and the metal
abundances of CRs, Be and B are produced at (above) the levels observed
in the most metal-poor stars with detection of Be or B, if the $^6$Li
plateau abundance is made by the same CCR nucleosynthesis.  The CR
acceleration energy needed to make $^6$Li primordial plateau abundance
at the observed level is $\sim 3-31$~\% of the SN kinetic energy.  This
value is not too large in view of an inferred present fraction of the SN energy
used to the CR acceleration~\citep{dru1989}.  The pregalactic SN
activity might have produced some level of light elements.  Although the
resulting abundances of light elements depend on the parameters which I
fixed to the values of \citet{dai2006}, the future measurements of
metal-poor stars would show the reasonableness of this early $^6$LiBeB
production mechanism, and might provide a signature of a primordial Be
(and perhaps B) plateau.  Further observations of LiBeB elements in
MPHSs are highly  desirable and valuable.

\acknowledgments

I thank Fr\'ed\'eric~Daigne for explaining the result of his chemical
evolution calculation, and Emmanuel~Rollinde for helpful comments.  I
am grateful to Adrian~Jenkins for giving me his code to make the mass
function of dark matter halos.  I appreciate the support of the Japan
Society for the Promotion of Science.

\clearpage

\begin{deluxetable}{ccccccc}
\tablecaption{Abundance results for two models\label{tab1}}
\tablewidth{0pt}
\tablehead{
\colhead{Model} & \colhead{$\epsilon$} & \colhead{$^{6}$Li/H} &
 \colhead{$^{7}$Li/H} & \colhead{$^{9}$Be/H} & \colhead{$^{10}$B/H} &
 \colhead{$^{11}$B/H}
}
\startdata
Model 1 & 0.31 & 6.0$\times~10^{-12}$ & 7.3$\times~10^{-12}$ & 9.0$\times~10^{-14}$ & 3.2$\times~10^{-13}$ & 7.6$\times~10^{-13}$ \\
Model 1 (Pop II only)& 0.73 & 6.0$\times~10^{-12}$ & 7.3$\times~10^{-12}$ & 5.7$\times~10^{-14}$ & 2.1$\times~10^{-13}$ & 4.9$\times~10^{-13}$ \\
Rapid burst model & 0.029 & 6.0$\times~10^{-12}$ & 7.9$\times~10^{-12}$ & 4.4$\times~10^{-13}$ & 2.0$\times~10^{-12}$ & 5.2$\times~10^{-12}$ \\
\enddata

\end{deluxetable}


\begin{thebibliography}{}
\bibitem[Asplund et al.(2006)]{asp2006} Asplund, M., Lambert, 
D.~L., Nissen, P.~E., Primas, F., \& Smith, V.~V.\ 2006, \apj, 644, 229 

\bibitem[Boesgaard et al.(1999)]{boe1999} Boesgaard, A.~M., 
Deliyannis, C.~P., King, J.~R., Ryan, S.~G., Vogt, S.~S., \& Beers, T.~C.\ 
1999, \aj, 117, 1549 

\bibitem[Boesgaard \& Novicki(2006)]{boe2006} Boesgaard, A.~M., 
\& Novicki, M.~C.\ 2006, \apj, 641, 1122 

\bibitem[Bond \& Efstathiou(1984)]{bon1984} Bond, J.~R., \& 
Efstathiou, G.\ 1984, \apjl, 285, L45 

\bibitem[Bonifacio et al.(2007)]{bon2007} Bonifacio, P., et al.\ 2007,
			      \aap, 462, 851 

\bibitem[Coc et al.(2004)]{coc2004} Coc, A., Vangioni-Flam, E., 
Descouvemont, P., Adahchour, A., \& Angulo, C.\ 2004, \apj, 600, 544 

\bibitem[Cumberbatch et al.(2007)]{cum2007} Cumberbatch, D., 
Ichikawa, K., Kawasaki, M., Kohri, K., Silk, J., 
\& Starkman, G.~D.\ 2007, \prd, 76, 123005 

\bibitem[Cunha et al.(2000)]{cun2000} Cunha, K., Smith, V.~V., 
Boesgaard, A.~M., \& Lambert, D.~L.\ 2000, \apj, 530, 939 

\bibitem[Daigne et al.(2006)]{dai2006} Daigne, F., Olive, 
K.~A., Silk, J., Stoehr, F., \& Vangioni, E.\ 2006, \apj, 647, 773 

\bibitem[Daigne et al.(2004)]{dai2004} Daigne, F., Olive, 
K.~A., Vangioni-Flam, E., Silk, J., \& Audouze, J.\ 2004, \apj, 617, 693 

\bibitem[Drury et al.(1989)]{dru1989} Drury, L.~O.,
			      Markiewicz, W.~J., \& Voelk, H.~J.\ 1989,
			      \aap, 225, 179

\bibitem[Duncan et al.(1997)]{dun1997} Duncan, D.~K., Primas, 
F., Rebull, L.~M., Boesgaard, A.~M., Deliyannis, C.~P., Hobbs, L.~M., King, 
J.~R., \& Ryan, S.~G.\ 1997, \apj, 488, 338 

\bibitem[Fields et al.(2000)]{fie2000} Fields, B.~D., Olive, 
K.~A., Vangioni-Flam, E., \& Cass{\'e}, M.\ 2000, \apj, 540, 930 

\bibitem[Garcia Lopez et al.(1998)]{gar1998} Garcia Lopez, 
R.~J., Lambert, D.~L., Edvardsson, B., Gustafsson, B., Kiselman, D., \& 
Rebolo, R.\ 1998, \apj, 500, 241 

\bibitem[Heger et al.(2003)]{heg2003} Heger, A., Fryer, C.~L., 
Woosley, S.~E., Langer, N., \& Hartmann, D.~H.\ 2003, \apj, 591, 288 

\bibitem[Heger \& Woosley(2002)]{heg2002} Heger, A., \& 
Woosley, S.~E.\ 2002, \apj, 567, 532 

\bibitem[Inoue et al.(2005)]{ino2005} Inoue, S., Aoki, W., 
Suzuki, T.~K., Kawanomoto, S., Garc{\'{\i}}a-P{\'e}rez, A.~E., Ryan, S.~G., 
\& Chiba, M.\ 2005, in IAU Symp. 228, From Lithium to Uranium: Elemental Tracers of Early 
Cosmic Evolution, ed. V.~Hill, P.~Fran{\c c}ois, \& F.~Primas (Cambridge: Cambridge University Press), 59 

\bibitem[Jedamzik(2000)]{jed2000} Jedamzik, K.\ 2000, \prl, 84, 3248 

\bibitem[Jedamzik(2004a)]{jed2004a} Jedamzik, K.\ 2004, \prd, 70, 
063524 

\bibitem[Jedamzik(2004b)]{jed2004b} Jedamzik, K.\ 2004, \prd, 70, 
08351

\bibitem[Jedamzik(2006)]{jed2006} Jedamzik, K.\ 2006, \prd, 74, 
103509 

\bibitem[Jenkins et al.(2001)]{jen2001} Jenkins, A., Frenk, 
C.~S., White, S.~D.~M., Colberg, J.~M., Cole, S., Evrard, A.~E., Couchman, 
H.~M.~P., \& Yoshida, N.\ 2001, \mnras, 321, 372 

\bibitem[Kawano(1992)]{kawano1992} Kawano, L.\ 1992, NASA 
STI/Recon Technical Report N, 92, 25163 

\bibitem[Kawasaki et al.(2005)]{kawasaki2005} Kawasaki, M., Kohri, 
K., \& Moroi, T.\ 2005, \prd, 71, 083502 

\bibitem[Kneller et al.(2003)]{kne2003} Kneller, J.~P., 
Phillips, J.~R., \& Walker, T.~P.\ 2003, \apj, 589, 217 

\bibitem[Kusakabe et al.(2006)]{kus2006} Kusakabe, M., Kajino, 
T., \& Mathews, G.~J.\ 2006, \prd, 74, 023526 

\bibitem[Letaw et al.(1983)]{let1983} Letaw, J.~R., Silberberg, 
R., \& Tsao, C.~H.\ 1983, \apjs, 51, 271 

\bibitem[Maeder \& Meynet(1989)]{mae1989} Maeder, A., \& 
Meynet, G.\ 1989, \aap, 210, 155 

\bibitem[Mathews et al.(2005)]{mat2005} Mathews, G.~J., Kajino, 
T., \& Shima, T.\ 2005, \prd, 71, 021302

\bibitem[Mel{\'e}ndez \& Ram{\'{\i}}rez(2004)]{mel2004} 
Mel{\'e}ndez, J., \& Ram{\'{\i}}rez, I.\ 2004, \apjl, 615, L33 

\bibitem[Meneguzzi et al.(1971)]{men1971} Meneguzzi, M., 
Audouze, J., \& Reeves, H.\ 1971, \aap, 15, 337 

\bibitem[Mercer et al.(2001)]{mer2001} Mercer, D.~J., et al.\ 
2001, \prc, 63, 065805 

\bibitem[Montmerle(1977)]{mon1977} Montmerle, T.\ 1977, \apj, 
216, 177

\bibitem[Piau et al.(2006)]{pia2006} Piau, L., Beers, T.~C., 
Balsara, D.~S., Sivarani, T., Truran, J.~W., 
\& Ferguson, J.~W.\ 2006, \apj, 653, 300 

\bibitem[Pospelov(2006)]{pos2006} Pospelov, M.\ 2006, preprint
			      (hep-ph/0605215) 

\bibitem[Prantzos(2006)]{pra2006} Prantzos, N.\ 2006, \aap,
			      448, 665 

\bibitem[Press \& Schechter(1974)]{pre1974} Press, W.~H., \& 
Schechter, P.\ 1974, \apj, 187, 425 

\bibitem[Primas et al.(2000a)]{pri2000a} Primas, F., Asplund, M., 
Nissen, P.~E., \& Hill, V.\ 2000, \aap, 364, L42 

\bibitem[Primas et al.(2000b)]{pri2000b} Primas, F., Molaro, P., 
Bonifacio, P., \& Hill, V.\ 2000, \aap, 362, 666 

\bibitem[Primas et al.(1999)]{pri1999} Primas, F., Duncan, 
D.~K., Peterson, R.~C., \& Thorburn, J.~A.\ 1999, \aap, 343, 545 

\bibitem[Ramaty et al.(1997)]{ram1997} Ramaty, R., Kozlovsky, 
B., Lingenfelter, R.~E., \& Reeves, H.\ 1997, \apj, 488, 730 

\bibitem[Read \& Viola(1984)]{rea1984} Read, S.~M., \& Viola, 
V.~E., Jr.\ 1984, Atomic Data and Nuclear Data Tables, 31, 359

\bibitem[Reeves(1974)]{ree1974} Reeves, R.\ 1974, \araa, 12, 
437 

\bibitem[Ryan et al.(2000)]{rya2000} Ryan, S.~G., Beers, T.~C., 
Olive, K.~A., Fields, B.~D., \& Norris, J.~E.\ 2000, \apjl, 530, L57 

\bibitem[Rollinde et al.(2005)]{rol2005} Rollinde, E., 
Vangioni, E., \& Olive, K.\ 2005, \apj, 627, 666 

\bibitem[Rollinde et al.(2006)]{rol2006} Rollinde, E., 
Vangioni, E., \& Olive, K.~A.\ 2006, \apj, 651, 658 

\bibitem[Rollinde et al.(2008)]{rol2008} Rollinde, E., Maurin, 
D., Vangioni, E., Olive, K.~A., \& Inoue, S.\ 2008, \apj, 673, 676 

\bibitem[Schaerer(2002)]{sch2002} Schaerer, D.\ 2002, \aap, 
382, 28

\bibitem[Sheth \& Tormen(1999)]{she1999} Sheth, R.~K., \& 
Tormen, G.\ 1999, \mnras, 308, 119 

\bibitem[Shi et al.(2007)]{shi2007} Shi, J.~R., Gehren, T., Zhang, H.~W.,
			      Zeng, J.~L., \& Zhao, G.\ 2007, \aap, 465,
			      587 

\bibitem[Spergel et al.(2003)]{spe2003} Spergel, D.~N., et al.\ 
2003, \apjs, 148, 175 

\bibitem[Spergel et al.(2007)]{spe2007} Spergel, D.~N., et al.\ 
2007, \apjs, 170, 377 

\bibitem[Spite \& Spite(1982)]{spi1982} Spite, F., \& Spite, 
M.\ 1982, \aap, 115, 357 

\bibitem[Suzuki \& Inoue(2002)]{suz2002} Suzuki, T.~K., \& 
Inoue, S.\ 2002, \apj, 573, 168 

\bibitem[Suzuki \& Yoshii(2001)]{suz2001} Suzuki, T.~K., \& 
Yoshii, Y.\ 2001, \apj, 549, 303 

\bibitem[Tatischeff \& Thibaud(2007)]{tat2007} Tatischeff, V., 
\& Thibaud, J.-P.\ 2007, \aap, 469, 265 

\bibitem[Valle et al.(2002)]{val2002} Valle, G., Ferrini, F., 
Galli, D., \& Shore, S.~N.\ 2002, \apj, 566, 252

\end{thebibliography}
\end{document}